\documentclass{ieeeaccess}
 
\usepackage{cite}
\usepackage{amsmath,amssymb,amsfonts}
\usepackage{graphicx}
\usepackage[hidelinks]{hyperref}
\usepackage{multirow}
\usepackage{xcolor}
\usepackage{colortbl}
\usepackage{enumitem}
\usepackage{dblfloatfix}
\usepackage{ulem}
\usepackage{hhline}
\usepackage{float}
\usepackage{algpseudocode}
\usepackage{algorithm}
\usepackage{framed}
\usepackage{adjustbox}
\def\BibTeX{{\rm B\kern-.05em{\sc i\kern-.025em b}\kern-.08em
    T\kern-.1667em\lower.7ex\hbox{E}\kern-.125emX}}

\newcommand{\github}{\url{https://github.com/zxcv123456qwe/iSNEAK}}
\newcommand{\IT}{\sffamily{iSNEAK}}

 \usepackage{mdframed}
 
\usepackage{enumitem}
 \setitemize{noitemsep,topsep=0pt,parsep=0pt,partopsep=0pt}
\newcommand{\bi}{\begin{itemize}[leftmargin=.5cm]}
\newcommand{\ei}{\end{itemize}}

\makeatletter
 {\par\unskip\endMakeFramed}
\makeatother

\definecolor{formalshade}{rgb}{0.93,0.93,0.93}

\definecolor{darkblue}{rgb}{0.2, 0.2, 0.2}

\newenvironment{formal}{%
  \def\FrameCommand{%
    \hspace{1pt}%
    {\color{darkblue}\vrule width 2pt}%
    {\color{formalshade}\vrule width 4pt}%
    \colorbox{formalshade}%
  }%
  \MakeFramed{\advance\hsize-\width\FrameRestore}%
  \noindent\hspace{-1pt}
}
{%
  \vspace{3pt}
  \endMakeFramed%
}
 
\begin{document}
\history{Date of publication xxxx 00, 0000, date of current version xxxx 00, 0000.}
\doi{}

\title{
iSNEAK: Partial Ordering as Heuristics
for Model-Based Reasoning in Software Engineering}
\author{
\uppercase{Andre~Lustosa}\authorrefmark{1},
\uppercase{Tim~Menzies}\authorrefmark{2},
}

\address[1]{North Carolina State University, Raleigh, NC  27695 USA (e-mail: alustos@ncsu.edu)}
\address[2]{North Carolina State University, Raleigh, NC 27695 USA (e-mail: timm@ieee.org)}

\tfootnote{...}

\markboth
{Author \headeretal: Preparation of Papers for IEEE TRANSACTIONS and JOURNALS}
{Author \headeretal: Preparation of Papers for IEEE TRANSACTIONS and JOURNALS}

\corresp{Corresponding author: Andre~Lustosa (e-mail: alustos@ncsu.edu).}

\begin{abstract}
A ``partial ordering'' is a way to heuristically order a set of examples (partial orderings
are a set where, for certain pairs of elements, one precedes
the other).  While these orderings may only be approximate,
they can be useful for guiding a search towards better regions of the data.
~\\
To illustrate the value of that technique, this paper presents
  {\IT}, an
incremental human-in-the-loop AI
problem  solver.
{\IT} uses partial orderings
and feedback from humans 
to prune the space of options.

Further, in experiments with a dozen software models of increasing size and complexity (with up to 10,000 variables),  {\IT}
 only asked a handful of questions to return human-acceptable solutions that outperformed the prior state-of-the-art.
 
We propose the use of partial orderings and tools like {\IT} to solve the {\bf information overload problem} where human experts grow fatigued and make mistakes when they are asked too many questions.   {\IT}  mitigates
  the information overload problem since it allows humans to explore
 complex problem spaces in far less time, with far less effort. 

 \end{abstract}

\begin{keywords}
Model-based reasoning, optimization, interactive search-based SE 
\end{keywords}

\titlepgskip=-21pt

\maketitle

\section{Introduction}\label{intro}

Model-based software engineering (MBSE) involves using high-level abstractions to guide discussions, code
generation, and validation~\cite{8804427}. In MBSE, models serve as blueprints for developers and automated tools to enhance productivity. For an example of model-based reasoning in software engineering, see our next section.

When faced with large and complex problems, humans use heuristic cues to lead them to the most important parts 
of a model~\cite{simon1956rational}.Such cues are essential if humans are to reason about large problems. But they can introduce their own errors:
 \begin{quote}
  ...people (including experts) are susceptible to ``automation bias'' (involving)  omission errors—failing to take action because the automated system did not provide an alert—and commission errors (where they incorrectly read model output)~\cite{green2022flaws}.
   \end{quote}
  This is troubling since there is an increasing
  legal demand for humans to certify that models are performing correctly~\cite{green2022flaws}. When humans are required
  to audit a model that they cannot understand,
  then this leads to  ``rubber stamping'' (model  approval  without proper consideration)
 and the   {\bf ``legitimizing the use of faulty and controversial (models) without addressing (their fundamental issues'')}~\cite{green2022flaws}. 

Accordingly,  we explore  {\bf interactive}  tools  
  where human knowledge and AI interact to find the factors that most change
  model conclusions.    Our {\IT} tool 
  asks humans the 
{\bf least questions} to find the {\bf most desirable solutions}.
Working with {\IT}, 
we have found 
that a small set of easily computed ``hints'' can generate an approximate 
  ``{\bf partial orderings}'' of the candidate solutions.  Using the orderings generated from those ``hints'', humans can guide a simple binary chop procedure over nested bi-clusters built in a space built through dimensionality reduction (see details, below). In this way, after offering $O(\log(N))$
opinions about $N$ examples, humans can find a small set of most acceptable solutions. 

Two  concerns with this procedure are:
\bi
\item
  {\bf AI inference can   confuse humans}; i.e. tools like
  {\IT}    would confuse people by either overwhelming
  people with too many questions or, at the end of the inference,
  propose a solution that made no sense to them.
 \item {\bf Humans' advice can
  confuse  AI  tools like {\IT}}; i.e. the guidance offered by
  humans would lead our AI algorithms to make sub-optimal
  conclusions. 
\ei
The rest of this paper   explores these concerns over  a set of a dozen models
of increasing size and complexity (with up to 10,000 variables).
These models address a range of problems including
software avionics;  flight guidance; management decisions
for agile software projects; and various software feature models including one electronic billing system.  

 Using these case studies, our experimental section explores
the two issues raised above:
\bi
\item
{\bf RQ1}:  Does {\IT}-style AI   confuse humans?
For our models, \underline{{\IT}-style AI does not confuse humans} since
it only   asks a handful of questions
and returns human-acceptable solutions; 
\item 
{\bf RQ2}: Does human advice confuse {\IT}-style AI?
For the models studied here, {\bf advice from humans do not confuse  AI} since {\IT}'s   solutions,
obtained via human-in-the-loop reasoning out-perform the prior state-of-the-art
in the optimization of software models. 
\ei 
We justify the use of partial orderings and tools like {\IT} to solve the {\bf information overload problem}. This problem is particularly acute in scenarios
 where humans are being asked their preferences for one
 solution over another. In that situation, information overload means   humans can make mistakes (as well as needing more
 time to accurately comment on examples).

 Overall, we say that  the contribution of  {\IT} is that it lets humans explore complex problem spaces within far less time, with far less effort, thereby avoiding information overload and its associated problems.  
To help with the reproduction and improvement of these results,  all our data and scripts are on-line\footnote{\github}.

The rest of this paper is structured based on advice by Wohlin \& Runeson et al.~\cite{wohlin2012experimentation} on empirical software engineering. We present our methods in Section \ref{isneak} and the comparable baselines in Section \ref{other}. Followed by a definition and explanation of our case studies in Section \ref{methods}, our results are in Section \ref{results}. See also Section \ref{threats} for a discussion on the threats to the validity of our conclusions.

\subsection{Current Limitations}\label{limit}
Before going on, we list some limitations to {\IT}.  
 One limitation of the current implementation is that it  assumes binary attributes (and n-valued attributes are handled with one-hot encoding based on equal-width binning). It would be useful to extend this code base to arbitrary
 precision numbers. Another more fundamental limitation is that in its current form, {\IT} is only defined for optimization problems, not
 regression, classification or generative tasks. We think more research can remove that limitation for regression tasks  (e.g. a regression problem can be viewed as an optimization problem where an optimizer explores different models to minimize  the difference between predicted and actuals). But as to classification and generative tasks, those are an open issue.

\newpage
\section{Background}\label{maths}
\subsection{Model-based Software Engineering is Widely Used }\label{mbse1}
 We study model-based methods since they are widely used in software engineering. For example,
 in the cyberphysical realm, it is standard
practice~\cite{8453177} to deliver hardware along with
a simulation system (often written in Simulink or C).
Developers use  that simulation  during verification. 

For another example, in 
 requirements engineering, 
 software product line researchers explore a space of inter-connected features in a software design~\cite{mendonca2009splot,89chen2018sampling,90nair2018finding,galindo2019automated,Kira92,hierons2016sip,el2014opti,saber2017seeding,pohl11}.  Such feature models
 can be reversed engineering from code, then used
 to debate proposed   changes
 to a system~\cite{sayyad2013optimum}.

Also, proponents of domain-specific languages
 argue for a systems-development life-cycle that begins with (a)~defining a domain language~\cite{fowler10_dsl},
 then (b)~modeling \& executing that system
 via that   language. Low code proponents
 then add some visual editor to allow
 for rapid changes to the model~\cite{prinz2021low}).

Further, software safety researchers generate   two models: one for the system under study and another for the safety properties.   Automatic tools then check for        violations of   safety  in the systems model~\cite{holzmann2014mars,8880058}.

Furthermore, models can be used for the generation of runtime
systems. This is especially useful for automotive
and embedded systems that make extensive use of models (e.g. generating executables from Simulink~\cite{mader2013oasis}.

 Lastly,
researchers in search-based
software engineering (SBSE) explore   models with sets of goals that may be  competing.
We have applied SBSE to models
of cloud-based architectures~\cite{chen2017riot}  (looking for ways to reduce the cost of the overall design);
NASA spaceships~\cite{DBLP:conf/re/FeatherM02} (looking for cheaper satellites that can handle more error conditions);
video encoders~\cite{nair18tse,peng2023veer} (looking for better and faster
compression control parameters);
software process models (looking for more functionality, at less cost) \cite{sayyad2013scalable}. 

 \subsection{Interactive Model-Based Software Engineering}\label{define}

\begin{figure*}[!t]
\hspace{2cm}\includegraphics[width=.8\linewidth]{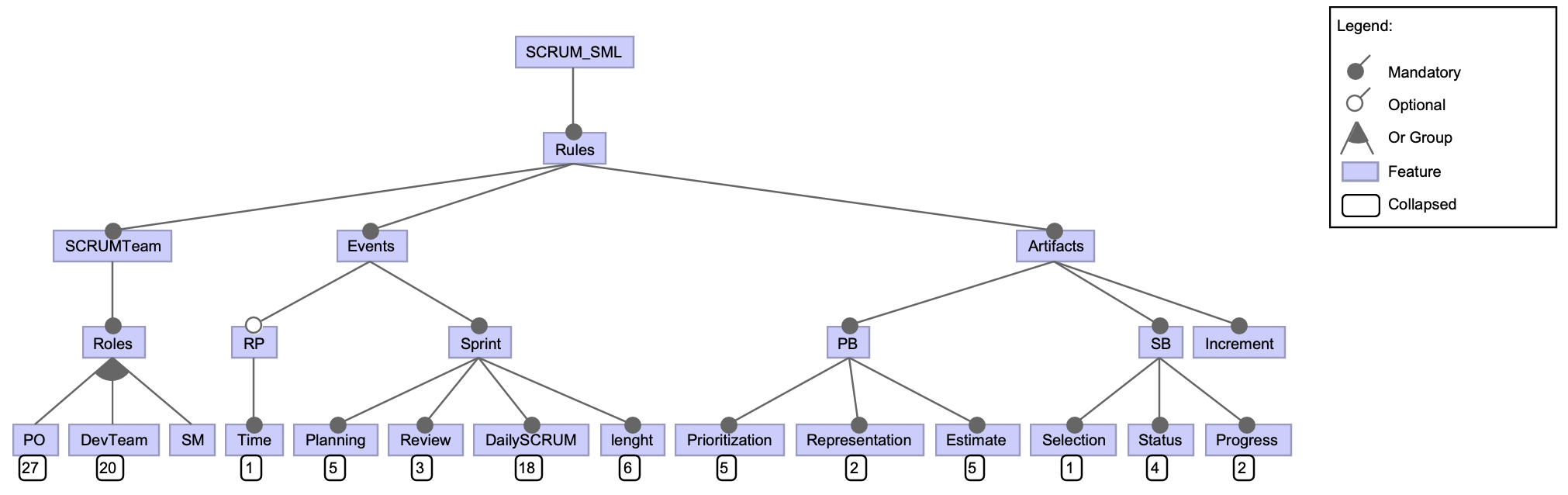}
  \caption{SCRUM Feature Model.
\textcolor{blue}{ Such feature models define features and their dependencies, typically in the form of a feature diagram + left-over (a.k.a. cross-tree) constraints.}
Numbers on leaves show the numbers of constraints in that part of the model. Not shown here, for space reasons, are   ``cross-tree constraints''
 that connect choices in different subtrees.}\label{fig:scrumModel}
\end{figure*}

Formally, then, we can define the problem we are trying to solve as such: 

Given a model M that contains $n$ variables, $c$ constraints and $g$ goals, how to find a {\bf valid candidate} solution that optimizes for the specific model M goals as well as maximizing user satisfaction (i.e: human input in what options they find important is being respected).

With the subgoal of minimizing the amount of human interaction ($I * S$) where $I$ is the number of times the algorithm goes to the user and $S$ is how much information is shown to the user to analyze at each given interaction

A {\bf valid candidate} solution is defined as a configuration of the given model that respects those model's constraints. When dealing with very large models with a high number of constraints, the space of candidate solutions is represented as all possible configurations of the model irrespective of whether they satisfy all constraints or not. This distinction is necessary because many of the methods studied in this arena use evolutionary computation algorithms, which are notoriously inefficient when dealing with highly constrained search spaces, those algorithms as we will show in our results section struggle with generating {\bf valid candidate} solutions, instead spending a large majority of their compute time reasoning over invalid space.

The reason for that is many of these models contain what is called a cross-tree constraint, which is a non-trivial constraint where there are conditions applied to configurations depending on different configurations somewhere else in the model. (e.g.: In the scrum model, if you choose for teams to have no Project Manager, you cannot then, choose for the Project Manager to lead the Daily meetings)

As with any optimization problem, a key concern is to report feasible solutions (where infeasible solutions violate domain constraints). A naive random model generation process is particularly prone to generating infeasible solutions. To solve that, we use a theorem prover to apply domain constraints
to find the feasible within the infeasible (see our use of PicoSAT in Section 2C). After that, all of the processing described in this paper occurs within the feasible space.

\subsection{Models Can Confuse People}\label{confuse}
 Having stated our goal, above, this goal lists some problems with achieving that goal.
 
This section argues that (c)~complex models can confuse and overwhelm stakeholders; and (b)~ standard AI
tools are not enough to address that confusion.

Figure~\ref{fig:scrumModel} is a visual representation of the 
Mendonca et al.~\cite{mendonca2009splot}'s software product line model of SCRUM (agile project development). 

The feature model of Figure~\ref{fig:scrumModel}
 contains 128 project management options and
   250+ constraints
 (e.g., if  sprints last two
weeks, then each individual task must take less than 10 days of programming).

We say that a {\bf candidate} is  one setting to all the variables in Figure~\ref{fig:scrumModel}. We also say that a {\bf valid candidate} is one such setting that respects all constraints present in the model.

We note that this is a multi-objective problem , meaning that all of the models in this study contain at least 4 goals. Each node in Figure~\ref{fig:scrumModel} comprises
(a)~an estimated development effort;
(b)~a purchasing cost (which is non-zero for features from   third-party libraries);
(c)~some number of defects seen in   past developments;
(d)~a ``success'' number that counts usage in prior successful projects. 
A project manager reviewing all the candidate solutions must reflect on how best to trade off between these competing concerns; e.g. how to   {\bf minimize} cost and effort without increasing defects
while   {\bf maximizing}
the features delivered. 

How to explore all the options within Figure~\ref{fig:scrumModel}? 
Empirically, it is known that if we select variable settings at random, then  less than 2\% of   those candidates
will satisfy the constraints of that figure~\cite{chen18}.  
 Modern AI tools like the  PicoSAT~\cite{biere2008picosat}  solver can navigate those constraints
 to find valid solutions\footnote{After one solution is found, its negation is added back to the model
and the SAT Solver is rerun to find a different solution.}.
But now there is a new problem: {\bf too many solutions}
(specifically  
$0.02*2^{128}\approx 10^{37}$ candidates).
To say the least, it is   hard   for humans to explore all   trade-offs while also satisfying   the
multiple model  constraints  within a space
 of   $10^{37}$  candidate solutions~\cite{lustosa21}.
So now we must ask:
\begin{formal}{\bf Q:}
 How to   help humans choose from  all  the candidates?
  \end{formal} 

\subsection{ Partial Orderings}\label{order}

This paper argues that an efficient way to choose between all the candidates   is {\bf partial orderings}.
A  partial order  is an ordering of a set such that, for certain pairs of elements, one precedes the other. 

 We use heuristic  partial orderings  as a way to quickly divide and conquer a problem space. For example, suppose there is a heuristic that Chinese manufacturing is better than Australian manufacturing. From this rule, we could define a partial ordering of the form:
\bi
\item Partially order anything from SE Asia from ``good'' to ``bad'' by mapping its manufacturing location (e.g. in the Philippines) to its closest
point on a line connecting Sydney, Australia to Shanghai, China. 
\item By this partial ordering, we might prefer to buy goods from Vietnam over goods from New Guinea. 
\ei
Like any heuristic, partial orderings are not a complete inference (since there may be pairs for which neither element precedes the other, according to  a partial ordering). However, as shown
below, they can be remarkably effective.

 A repeated result in software engineering (and other domains)~\cite{ratner2019training, Pornprasit23, wang2019characterizing, nair2017using} is that when optimizing for some goals, it is possible to be guided by some easy-to-compute partial  heuristic.
 For example, 
automatic configuration assessment tools can be built from regression trees with just a few examples. Even if those trees become very poor predictors of performance (in an absolute sense), they can still be useful to rank (in a relative sense) different configurations.
E.g.,  Nair, et al. used those trees to find the top 1\% of  configurations,  even though those models   had error rates as high as
90\%~\cite{nair2017using}.  
 Also, some researchers~\cite{Pornprasit23} use generative models to improve upon supervised learning. While the results from the generative model may not be of the highest quality, all these outputs provide hints on how to better direct another algorithm (e.g., a machine learner).
Further, researchers explored test case selection via crowd-sourcing since this is a fast way to collect many  opinions. While the value of one opinion  is questionable, they can be useful in the aggregate~\cite{wang2019characterizing}.  
 
To say all this another way, SE researchers often   guide the search for ``good'' solutions via some heuristic that can pre-order the space of possible solutions. To demonstrate the power of this approach, we turn to Hamlet's probable correctness theory~\cite{hamlet1987probable}. That  theory says that the
confidence of  seeing an event occurring at probability $p$ after $n$ trials is $c=1 - (1-p)^n$, which, if rearranged, tells us how many samples are required to find something: 
\begin{equation}
n(c,p)=\frac{log(1-c)}{log(1-p)}
\end{equation}
For example, for 99.9\% confidence  of seeing the top 1\% of all solutions, humans must  evaluate $\approx 688$ candidates.

Now consider what happens if we can compute some partial ordering over all those solutions.  
For solutions sorted in that manner,
 Hoare's quickselect algorithm\footnote{
Quickselect uses a similar approach to quicksort (choosing one item as a pivot and partitioning the data in two, based on a said pivot), but instead of recursing on both sides,  quickselect only recurses on the better  side~\cite{hoare61}.}   needs only 
explore:
\begin{equation}\label{eqeval}\begin{array}{rcl}
\mathit{evaluations}&= &2\log_2(n)
\end{array} 
\end{equation}
of the above $n$ samples\footnote{Why do we multiply by 2 in Eq.~\ref{eqeval}? As discussed later in this paper,
our algorithm evaluates one candidate from each sibling subtree: see \ref{isneak} for details.}.
Assuming this binary chop, and repeating the above calculation,  the equations say that now we only need to evaluate $\log_2(n=688)\approx 10$ times.
If the reader finds this calculation to be over-optimistic, we point out that
our experimental results in \ref{rq2} show that  {\IT} can find
good candidate solutions after  just a few   dozen evaluations.

One way to summarize this equation is to say  that optimization problems can be easier than  other tasks like classification or regression. Those other tasks
require us to label examples across the entire space while
optimization means looking for a small region (and pruning the rest). And that small region can be found via partial ordering.

When discussing this work, we are often asked how 
this work compares
to other SE researchers in what they call  ``partial orderings''.
For example, researchers trying to reduce   the state space explosion of model 
 checking~\cite{peled1998ten}, or simplifying the  number of refactoring
 steps for source code~\cite{MORALES201825}, apply a technique they
 call {\bf partial order reduction} that prunes  similar paths leading to the same result. 
 While that work certainly shares some of the intuitions of this work, we note
 that (a)~our partial orderings sort the entire space of solutions; after which we
 (b)~explore $N$ solutions using $O(\log(N))$ evaluations. 
 In this second step, what is important in our work is that
 our use of partial orderings   lets humans guide the algorithms to their conclusions.

 \subsection{Why Not Explore Models Using Traditional Optimization Methods?}\label{defineRQs}

 This paper is about optimization. Optimization problems, like those discussed above, have been well-studied for decades
in the operations research literature~\cite{winston2022operations}.  
There are many reasons to be cautious about using  traditional optimizers for model-based SE.
Those optimizers  often assume continuous   functions   with numeric attributes that   can be differentiated at all points along a function (differentiability is a key requirement for any classical gradient descent algorithm). Models like Figure~\ref{fig:scrumModel}, on the other hand, have many non-numeric attributes and are not expressed in the equational form needed by traditional optimizers.

Also, traditional optimizers 
may make unrealistic assumptions about the nature of model goals.
Researchers like Harman~\cite{IEEE:Harman}  note that SE often requires trading off between competing 
many goals and, in that context, there may not be a single  optimal but rather a (small) space of most acceptable solutions which must be debated by stakeholders.  

Further, if used ``off-the-shelf'', traditional optimizers can waste much
time exploring issues that many humans find irrelevant or incomprehensible. 
While models may contain many variables, any one  subject matter expert may only have experience with a small number of them.  One result from the user study
of this paper is that when stakeholders are asked to comment on all model attributes (e.g., the 128 attributes of Figure~\ref{fig:scrumModel}), they only ever comment on around 20  attributes. 
This means that, in our experiments
on Figure~\ref{fig:scrumModel},
the space where their expertise
can be applied is just  $\ 2^{20}/2^{128} \approx  10^{-33}$
of the total state space. 
Hence we seek methods that can respond well to  very small samples
of human insight. 

\subsection{Reasoning over Many Goals}\label{other}

Many    rival  technologies that 
have been applied to many-goal model-based reasoning in SE:
\bi 
\item
Genetic algorithms, used for interactive search-based
SE;
\item  Sequential model optimizers, used for SE configuration. 
\ei
 This section describes those two approaches.
 
\subsubsection{SBSE and Interactive search-based
SE}\label{isbse}

Search-based SE methods explore design
trade-offs using
 evolutionary programs. These algorithms
 run  in multiple generations $G_i$:
\begin{enumerate}
\item Generation $G_0$ is the initial population.
\item Each new generation $G_{i+1}$ is built by 
\begin{enumerate}
    \item 
Randomly {\bf mutating}
the   $G_i$ solutions,
\item
Then {\bf selecting} the best individuals,
\item {\bf y-evaluating} these individuals
\item
Then mixing together (also known as  {\bf crossover})
parts of two best individuals to create a new example for $G_{i+1}$
\end{enumerate}
\end{enumerate}
This kind of reasoning  must  assess and compare solutions (in step~2b of the above). In single-objective optimization problems, a simple sorting function can rank goals between candidates. However, when dealing with many-objective reasoning, candidates must be ranked across many goals. 
As the number of goals increases,
simple schemes such as {\bf boolean domination}
find it harder to distinguish different candidates\footnote{Given two candidates with $n$ goals, one is better than the other if
(a)~none of its goals are  worse than the other, and (b)~at least one
goal is better. As the number of goals increases, it becomes increasingly
likely that at least one goal is worse, even if only by a small amount.
Hence researchers like Sayyad et al.\cite{sayyad2013value}
and Wagner et al.~\cite{Wagner07} deprecate boolean domination
for three or more goals. }. Hence, we use the Zitler predicate described in
Equation~\ref{eqzitzler}.

Interactive SBSE  (iSBSE) is a variant of SBSE that tries to include humans in the reasoning process.  
One drawback with these iSBSE  tools is {\bf cognitive fatigue}. Typical control policies for
genetic algorithms are to 100 individuals in $G_0$, which are then mutated for a hundred generations~\cite{holland92}. Most stakeholders can only accurately  evaluate a small fraction of the 100*100=10,000 individuals generated in this way.

To reduce cognitive fatigue, iSBSE researchers augment  
  genetic algorithms with   tools that enable pruning
  the search space, without having to ask the stakeholders too many  questions.
For example,  Palma et al. \cite{palma2011using} use a   constraint solver
(the MAX-SMT algorithm)  
to  evaluate  pair-wise comparisons of partial model solutions to decrease their search space.
We do not compare {\IT}
against this method since their technique has scaling problems:
\bi 
\item
Their biggest model had   50 variable   constraints.
\item {\IT}, on the other hand, has been successfully
applied to a 1000 variable model: see \ref{rq2}.
\ei
In other work, 
Lin et al.' \cite{lin2016interactive}'s iSBSE
tool generates plans of what to change
within a system.   Then,
stakeholders
interactively examines the recommended steps to accept, reject, or ignore them. These interactions will then be used as feedback to calculate the next recommendation.  But   like Palma et al.,  Lin et al. 
did not demonstrate that their methods
can scale to the same size models as we process later
in this paper.

In yet other work,  Araugo et al. \cite{araujo2017architecture}   combine
an interactive genetic algorithm with a machine learner. Initially, humans are utilized
to evaluate examples, but once there are enough examples to train a 
{\bf surrogate}; i.e., a model learned via machine
learner that can comment on solutions, this surrogate starts answering queries without
having to trouble the stakeholders.  

  In the future work section of Araújo et al.,
  they recommended exploring   software product lines 
such as
Figure~\ref{fig:scrumModel}. Since they have approved
that kind of study, our study  compares {\IT} against      Araújo et al. (as well as SMBO,  described below).

\begin{table}[!b]\small
\begin{mdframed}[backgroundcolor=blue!3,rightline=false,leftline=false] 
\small
\setlength{\leftmargini}{10pt}
\begin{enumerate}
\item {\bf SAMPLE}: From $N$ examples,  a small sample of  $m$ candidate solutions are y-evaluated. 
\item {\bf GENERALIZE}: Using the sample of $m$ y-evaluated solutions,  a surrogate is built via machine learning.
\item {\bf EXPLORE}: The surrogate  make guesses about the remaining 
\mbox{$N-m$} number  of non y-evaluated candidates. 
\item {\bf ACQUIRE}: The ``most interesting'' guess
(as determined by an {\bf acquisition function} is then y-evaluated.
\item {\bf LOOP}: $m=m+1$; Exit if $m>\mathit{budget}$. Else goto step 2.
\end{enumerate}
\end{mdframed}
\caption{Sequential Model-Based Optimization:  general framework.
 Step 4 of  Table~\ref{flash} shows a  specific acquisition function. }\label{smbo}
\end{table}

\subsubsection{SMBO = Sequential Model-Based Optimization}\label{smo}

Recall that  Araújo et al. reduced the number of questions to
the stakeholders by building a surrogate. Once that surrogate
is available then
 when new solutions
are generated, these are evaluated by the surrogate
(without having to ask the stakeholders any questions).

\begin{table}[!t]
\small
\begin{mdframed}[backgroundcolor=blue!3,rightline=false,leftline=false] 
\small
\setlength{\leftmargini}{10pt}
 \begin{enumerate}
\item {\bf SAMPLE}:  From $N$ examples, a  sample of $m_0=60$ candidate solutions are y-evaluated.
\item {\bf GENERALIZE}: Using the sample of $m$ evaluated solutions,  a regression tree is built by CART~\cite{BreiFrieStonOlsh84}, one tree per model objective.
\item {\bf EXPLORE}: These CART models  make guesses about the remaining $N-m$  non y-evaluated candidates. 
\item {\bf ACQUIRE}: In FLASH, the  ``most interesting'' guess
is the one that  has  maximum objective values across $n$ random projections.
\begin{enumerate}
\item Each non-y-evaluated candidate has $G$ goals $(g_1,g_2..,g_G)$   (computed by   the CART model learned in step 2). FLASH normalizes
all these goals 0..1 for min..max.
\item Each  non-y-evaluated candidate has $G$ weights \mbox{$(w_1,w_2,...w_G)$}
where $w_i=(-1,1)$ for goals being minimized or maximized (respectively); 
\item Each non-y-evaluated candidate's goals is   multiplied by a set of $G$  random weights $ (r_1,r_2,..r_G)$
where $0 \le r_i \le 1$.
\item The most interesting non y-evaluated candidate has max $\frac{1}{N-m}\sum_i^G g_i{\times}w_i{\times}r_i$
\end{enumerate}
\item {\bf LOOP}: $m=m+1$; Exit if $m>(\mathit{budget}=120)$. Else  go to step 2.
\end{enumerate}
\end{mdframed}
\caption{A variant of SMBO, as seen in   FLASH~\cite{flash_vivek}.
FLASH's use of random projections (in step 4c) was inspired
by earlier work on the   MOEA/D algorithm~\cite{zhang2007moea}.
The control parameters shown in this figure ($m_0=60,\mathit{budget}=120$) come from 
prior work~\cite{flash_vivek}.  }\label{flash}
\end{table}

The use of surrogates is an important technique  in the {\bf automatic configuration}
literature. Traditional configuration methods require the evaluation of many candidates. 
For example, in 2013, Apel et al. proposed the use of the CART regression tree learner 
to generate a model from some historical examples that could be used to assess configuration options for future configuration problems~\cite{Guo13}. That said, literature reviews in this field such as \cite{ochoa2018systematic} lament the narrow range   
of algorithms used in this area. Firstly many of these tools still require a large number of pre-evaluated candidates. Secondly,  the  configuration tools reported by  Ochoa et.al.
(that reportedly  supposedly supports human interaction),  
often use a human ranking of attributes that is fixed for all candidates~\cite{ochoa2018systematic}. Hence those reportedly interactive tools can overlook
     nuances related to  specific examples.

Nair et al. use  surrogates in a different way. In their tool, new examples
are assessed on a case-by-case basis by    {\bf sequential model-based optimization}~\cite{hutter2011sequential,zuluaga2013active} (SMBO).
Table~\ref{smbo} shows their {\bf acquisition function} loop that
uses the surrogate to 
guess which candidate  should be reviewed next.

Acquisition functions are the core of tools that use SMBO.
While working on automatic software configuration, 
Nair et al.   explored several
acquisition functions~\cite{flash_vivek}.
For SE models, they found that  standard SMBO methods
(using Gaussian Process Models) did not scale very well.
Instead, they found that
regression tree models based on CART~\cite{BreiFrieStonOlsh84}
scaled up to the kinds of large models seen in SE.
Following  their recommendations, this paper will use their 
FLASH   sequential model optimizer,  described
in Table~\ref{flash}. Note step 4d, where one new example is evaluated. This is the point
where some oracle would be asked for their opinion on a candidate (e.g., some human could be asked a question or some model could be executed to obtain y-values).

\section{ Our Proposal: The {\IT}  Algorithm}\label{isneak}
{\IT} uses   ``hints'' from the independent variables to   quickly generate a partial ordering of 
candidates. This space is then explored via  a recursive binary chop algorithm
that spaces the candidates  over  a PCA-like space.

\subsection{Recursive Bi-Clustering}
{\IT}  uses a recursive binary chop
inspired by 
Chen~\cite{chen2018sampling}.
This approach finds two very distant points {\bf east,west}, then recurses on the half
associated with the ``best'' of {\bf east} or {\bf west}
(where ``best'' is calculated using the methods described below).  Formally, this process is a recursive FASTMAP algorithm~\cite{faloutsos95fastmap} which is   Nystr\"om  ~\cite{platt04} approximation to PCA. We provide a pseudocode for FASTMAP in Alg.~\ref{fastmappseudo}.  The recursive version of this procedure simply consists on calling fastmap again on each of the (east and west) generated clusters from the latest step. This terminates on a given stopping criteria.

{\IT}'s recursion  terminates when   it reaches $n=\sqrt{N}$ examples. This stopping threshold was selected based on advice from Webb et al.~\cite{webb09} on how best to divide data.

\begin{algorithm}
\caption{FASTMAP}
\footnotesize
\begin{algorithmic}
\Procedure{FASTMAP}{rows}
  \State pivot = random(rows)
  \State east = mostDistant(pivot)
  \State west = mostDistant(east)
  \State projectedPoints = []
  \For{\texttt{row in rows}}
    \State project row in the line between east and west
    \State projectedRows.append(projectedRow)
  \EndFor
  \State split projectedRows in half
  \State first half is EastRows
  \State second half is WestRows
  \State \textbf{Return} EastRows, WestRows
\EndProcedure
\end{algorithmic}
\label{fastmappseudo}
\end{algorithm}

\subsection{Discretization}\label{bins}

This algorithm uses discretized attributes.
Following the advice of  Kerber~\cite{kerber1992chimerge}, numbers are initially divided into 16 ranges of equal-width bins. 
This division (into many small bins) is then carefully evaluated within
our system and (sometimes) repaired.
Specifically, our system checks if two adjacent bins can be combined 
to use statistical measures of   disorder.
To implement this, we find the size $n_i$ and standard deviation 
$\sigma_i$ of two adjacent bins $i,j$, then compare them with a bin $k$
formed by combining $i,j$. Our system combines bins if:
\[
\sigma_{k} \leq \frac{n_i}{n_k}\sigma_{i} + \frac{n_j}{n_k}\sigma_{j}\]

In early experimental work on this paper, we have attempted to fixate different values for the initial number of bins. But the results were similar or worse. As such we have kept the initial division of 16 ranges by Kerber~\cite{kerber1992chimerge}.

\subsection{Distance,   Projection, and Division}

For the descretized  data represented as one-hot encoding, 
Chen et al.'s boolean distance metric~\cite{chen2018sampling} is used to measure  the distance between two candidates. 
With that distance metric, the two remote points {\bf east,west} can be found in linear time, using the Faloutsos~\cite{faloutsos95fastmap} heuristic\footnote{Pick any point $X$ at random. Let {\bf east} the point furthest from $X$. Let {\bf west} be the candidate furthest from {\bf east}. For $N$ examples, this heuristic needs  $2N$
distance calculations (and a complete search would need  $N^2$ calculations).
} and the cosine rule:

\bi
\item Let $E$  ({\bf east}) and 
  $W$ ({\bf west}) be separated by distance $c$.
  \item Let $a,b$ be the distance of some example to $E,W$.
  \item
    All   examples have distances $a,b$  to $E,W$  and distance
    \mbox{$x=(a^2+c^2-b^2) / (2ac)$} on a line drawn $E$ to $W$.
    \item
    To halve the data, {\IT} splits   on the median $x$ value.
    \ei

\subsection{Ranking Best and Rest}\label{bore}
At each level of the recursion, 
the division process of the last section generates two halves, which {\IT} can rank  in one of two ways:
(a)~automatically; or
(b)~human-in-the-loop or 
Once ranked,
the lower ranked half is  pruned and   {\IT} recurses into the surviving   candidates.

\subsubsection{Automatic Ranking}\label{bore} Following the advice of 
Sayyad et al.\cite{sayyad2013value} {\IT} uses Zitzler's {\bf continuous domination predicate}~\cite{zitzler2004indicator} to label our two halves {\bf best} and {\bf rest}.
 This  predicate   favors  
$E$ over $W$   if jumping from $E$ ``loses'' most:
\begin{equation}\label{eqzitzler}\footnotesize
\begin{array}{rcl}
        \textit{worse}(E,W) &=&\textit{loss}(E,W) > \textit{loss}(W,E)  \\
        \textit{loss}(x,y) &=&\sum_{j=1}^n -e^{\Delta(j,x,y,n)}    \\
        \Delta(j,x,y,n) &=&w_j(o(x)_j  -  o(y)_j) / n
   \end{array}
   \end{equation}
where ``$n$'' is the number of objectives; $w_j\in \{-1,1\}$ depending on if we are maximizing goal $x_j$. $o$  are the raw objective scores (normalized 0..1 for min to max).

\subsubsection{Human-in-the loop Ranking}\label{human} {\IT} can rank the two halves using human input to label our two halves {\bf best} and {\bf rest}.
Starting with two items  picked  from each half, 
 {\IT} looks at the top
six more informative attributes (where ``most informative'' is assessed via the
INFOGAIN feature ranking predicate~\cite{Hall03}) that have different values in item1 and item2.
These two sets (of   up to six values) are   presented to the user. The user then selects either item1 values or item1 values.
Each half can then be scored by the percentage $P$ of candidates
in that half 
  containing any of the selected values.  This frequency then becomes an additional term in a separate version of  
Equation~\ref{eqzitzler}: 
\begin{equation}\label{eqzitzler1}
\footnotesize
\begin{array}{rcl}
\textit{worse}(E,W) &=&\textit{loss}(E,W) > \textit{loss}(W,E)  \\
        \textit{loss}(x,y) &=&\left(\sum_{j=1}^n -e^{\Delta(j,x,y,n+1)}\right)     
        -e^{\Delta'(x,y,n+1)}\\
         \Delta(j,x,y,n) &=&w_j(o(x)_j  -  o(y)_j) / n\\
          \Delta'(x,y,n) &=&(P(x)  -  P(y)) / n
   \end{array}
   \end{equation}

\begin{figure}[t!]
    \centering    
    \includegraphics[width=.8\columnwidth]{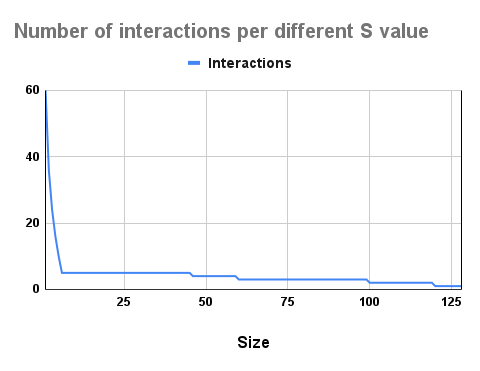}
    
    \caption{Experimental results from altering the {\bf size} of each interation on the Scrum model of Figure~\ref{fig:scrumModel}}
    \label{fig:IvsS}
\end{figure}

The reasoning behind showing 6 attributes to the user comes from experimental data. In the field of Interactive Search Based Software Engineering, a big concern is to avoid information overload. Using the example model from Figure~\ref{fig:scrumModel}, even if humans felt they could comment on all those 128 options at once, should we trust their assessment? Shackleford et al.~\cite{shackelford2007implementation} warns us that human information overload leads to errors in human decisions about which variables are most influential in a large comparison scenario. And Takagi~\cite{takagi1998interactive, takagi2000active, takagi2001interactive} notes that we can decrease this effect by:
\bi
\item Reducing $I$, the number of {\bf interactions} (where at each interaction, we showcase the user N attributes)
\item Reducing  $S$ the {\bf size} of questions per interaction
\ei
We note that:
\bi 
\item 
Every run of {\IT} could use a single interaction $I=1$ if 
an expert commented  on all the $n$ variables in our models (if $S=n$).
\item 
Similarly, {\IT} has  (at most) $n$ interactions if every interaction only commented on one variable
(if $S=1$).
\ei
Figure~\ref{fig:IvsS} explores the trade-off between asking for everything ($S=n$), or asking for one thing 
($S=1$) for the SCRUM model. We note that there is a non-linear relationship between $S$ and $I$. Specifically, there is a ``knee''
in this curve at $S=6$, below which the number of interactions spike (to some very large values). To avoid this
spike, we run our experiments at $S=6$.  As to the merits of other $S$ values,
we leave that for further work.

\subsection{Two Passes}
 Our  {\IT} system is a two-pass system. 
 Both passes apply recursive bi-clustering
 with top-down pruning. 
 In the second pass, 
 Chen's algorithm~\cite{chen2018sampling}   prunes
 subtrees via the automatic
 ranking method of \ref{bore}
 and Equation~\ref{eqzitzler}.  
 
Before that,
{\IT}'s pass1    uses 
 Equation~\ref{eqzitzler1} to guide
 the search via human preferences. Pass1 recursively bi-clusters all the candidates (with no pruning). Next, we generate
 a list of subtrees in the cluster tree,
 sorted by how much that subtree reduces
 the variability in the candidates.
Here, variability is measured via entropy.
If $e_0,e_1,e_2$ is the entropy of the independent attributes in a subtree
parent and its two subtrees, then 
  Equation~\ref{eqzitzler1} is applied
  to the subtree that maximizes 
\begin{equation}\label{open}
S{\times}(e_0-e_1){\times}(e_0-e_2)\frac{{\mathit open}}{d}
\end{equation}
where $d$ is the depth  of the root of this subtree\footnote{By definition,  the  root node of the cluster tree
 has depth  $d=1$.}.
  The $d$ term in Equation~\ref{open} means that we favor pruning larger subtrees (i.e. those closest to the root of the tree).
  
The {\bf open} term in  Equation~\ref{open}    constrains the dialog with users
  such that they rarely have to offer an opinion about the same attribute twice.   Once \ref{human} determines what attributes  to we could ask
  about this subtree, {\bf open} is percentage of those attributes
that we have {\bf not} yet presented to   users (so the subtrees we explore are the ones converned with attributes we have yet to ask about).   

 Pass1   continues until
    the cluster tree contains less than $\sqrt{N}$
  candidates (or every attribute has been asked, so Equation~\ref{open} returns all zeros).
   These survivors are then given to pass2 which returns   $\sqrt{\sqrt{N}}$ candidates. 
This two pass system can be seen in a summarized Pseudocode in Algorithm~\ref{sneakpseudo}.

\begin{algorithm}
\caption{\IT}
\footnotesize
\begin{algorithmic}
\Procedure{\IT}{rows}
  \State tree = recursiveFASTMAP(rows)
  \State good = calculateGood(tree)
  \If{good and size(tree) $\leq$ stop}
    \State Ask question on best
    \State prune worst half
    \State recalculate good
  \Else
    \State survivors = recoverRows(tree)
    \State selected = SWAY(survivors)~\cite{chen2018sampling}
    \State return selected
  \EndIf
\EndProcedure
\end{algorithmic}
\label{sneakpseudo}
\end{algorithm}

\begin{table}[!t]
\footnotesize
\begin{center}
\begin{tabular}{|r|clc|}\cline{2-4}
 \multicolumn{1}{c|}{~}         & ``Free'' variables                     & Constraint& Number\\ 
  \multicolumn{1}{c|}{~}        & (i.e., values that can be adjusted). & ratio& of goals\\\hline
OSP2  &   ~~~   6                                    &0   & 5\\
POM3A  &  ~~~  9                                     & 0   & 4\\
POM3B  &  ~~~  9                                     & 0   & 4\\
POM3C  & ~~~   9                                     & 0   & 4\\
FLIGHT  & ~~  11                                     & 0   & 5\\
GROUND  & ~~  12                                     & 0  & 5\\
BILLING  & ~~  88                                    & 1.02 & 4\\
125FEAT  & ~ 125                                     & 0.25  & 4\\
SCRUM  &  ~ 128                                      &  0.97 & 4\\
250FEAT &    ~ 250                                   &  0.25  & 4\\
.25 C.D.  &            ~ 500                            &0.25  & 4\\
.50 C.D.  &           ~  500                            & 0.50  & 4\\
.75 C.D.  &          ~   500                          &0.75  & 4\\
1.00 C.D.  &       ~ 500                                 & 1.00  & 4\\
500FEAT  &         ~ 500                                &   0.25& 4\\
1000FEAT  &    1000                                     &  0.25& 4\\ \hline
\end{tabular}
\end{center}
\caption{Number of variables being optimized and \underline{constraint ratio}.
For the   models on the first few rows, the 
\underline{constraint ratio} is zero (since every solution from those models
is valid). For the other models, the
\underline{constraint ratio} is the 
ratio  of constraints per clause.}\label{compare}
\end{table}

\section{Methods}\label{methods}

\subsection{Case Studies}
 
Table~\ref{compare} shows the models used in our study.
For all the models, the experiments of this paper (see next section)
apply the methods of \ref{isneak} and \ref{other} to find 
inputs that leads to best values in  outputs.

We use these models since
these are standard models used in other SBSE papers~\cite{chen2018sampling, menzies2006recognizing, chen2016evosampling}. Note that these are all
multi-objective problems with 4 to 5 goals. In that table,
the {\bf constraint ratio} is the 
number  of constraints per clause.
The models {\bf POM3A, POM3B, POM3C, OSP2, FLIGHT and GROUND}
are small and have no constraints (so all solutions generated to these models are valid).
All the other models are much larger and have many constraints
that only   1\% to 3\%  of the solution space is valid.

The rest of this section describes these case studies.

\subsubsection{XOMO}

{OSP2, GROUND and FLIGHT are variants of XOMO~\cite{menzies2005xomo} which combine four different
COCOMO-like software process models in order to calculate   metrics for   project's success. 
The XOMO model, is a general framework for Monte Carlo simulations that combine four COCOMO-like
software process models from Boehm's group at the University of California. 

The problem solved by {\bf XOMO}+{\IT} is to ``generate good management recommendations'' that satisfy the constraints of that project, while also minimizing   the development time (as judged by the COCOMO model) while also avoiding  other factors that
increase the risk of cost overrun   (as judged by other Boehm models).

Containing between 6 and 12 variables on each of its variants, the XOMO model is a 
representation of real situations in a software project. Under this model, a user can obtain four 
objective scores: (1) project risk; (2) development effort; (3) predicted defects; (4) total time for development.
The model was developed with data collected from hundreds of commercial and Defense Department projects \cite{boehm2003using}.
As to its risk model, it is defined as a rule-based algorithm in certain variables on these models associated with risk
(e.g.: demanding more reliability whilst decreasing analyst capability). As such XOMO measures risk as the percentage of triggered rules

{\bf  OSP2}   (short for ``orbital space plane, version2'') describes the software
 context of a second-generation NASA space    plane. Expressed in the terminology of Boehm's COCOMO system~\cite{boehm1995cost}, {\bf  OSP2} lists the range of acceptable
software reliability (it must be very high), database size (which is variable), developer experience (which can be changed by management assigning difference developers to the project), and so on. 

~{\bf GROUND, FLIGHT} are similar to {\bf OSP2}, but represent two specialized classes of NASA software (flight software and ground software). The  problem solved  by these  two models are the same   as {\bf OSP2} (but for special   kind of specific software types).

\begin{figure}[!t]
\includegraphics[width=\linewidth]{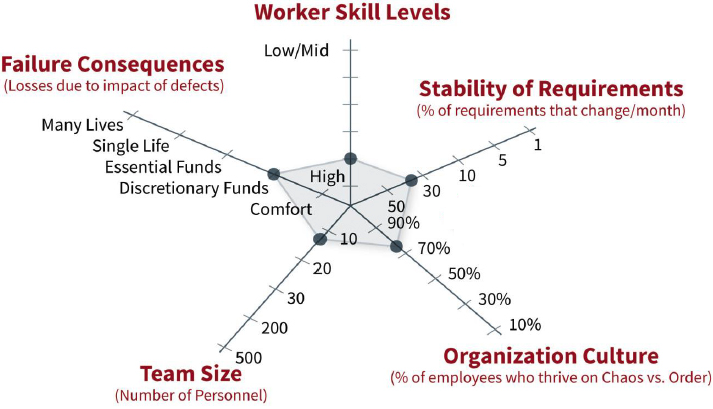}
 \caption{According to Boehm and Turner~\cite{boehm2003balancing},  agile is best suited for projects
in the middle of this figure.}\label{agdiag}
 \end{figure}

\subsubsection{POM3}
{\bf OSP2,GROUND,FLIGHT} assume waterfall style development.
{\bf POM3}, on the other hand, explores factors related to agile development such as Figure \ref{agdiag}.  These   factors are organization factors identified by  Boehm and Turner's in their  book: ``Balancing Agility and Discipline: A Guide for the Perplexed''~\cite{boehm2003balancing}. According to
that book,  agile works best in the center of Figure \ref{agdiag}. However,
they also note that with management support, organizations can adjust themselves.

{\bf  POM3}+{\IT} is an advice tool  used during SCRUM meetings
to answer the question ``what    tasks to complete in the next sprint?'' (where a sprint is a very short period of intense coding with the aim
 to produce some specific deliverables). To answer that question, POM3 conducts trade-off
 studies within   Figure \ref{agdiag}.

This paper explores three POM variants: 
\bi\item
{\bf POM3a} representing a very broad space of
projects; and 
\item 
{\bf POM3b} representing safety critical small  projects;  and 
\item 
{\bf POM3c} representing  highly dynamic
large projects.
\ei

  For all these {\bf POM}  models, the goal is to maximize the number of requirements met while minimizing the time to completion and idle time (time spent waiting for some other team to complete some required module).

  \subsubsection{Assorted Product Lines}

 The rest of our models come from the software product line research. Here,  a large space of potential software projects are modeled as a tree of features (and super-features are meet via some conjunction/disjunction of subtrees).  Between the tree, there may be ``cross-tree constraints''; i.e.
 things that must be absent or present in one branch to enable a feature in another branch. From one product line,
 many software designs can be generated, provided that each satisfy the cross-tree constraints. 
For an example software product line, see Figure~\ref{fig:scrumModel}. 
 
 When combined with some optimizing technology like {\IT}, the  problem solved  by these models plus  {\IT}
 is ``what product to build?''.
 To answer that question, an optimizer must trade off between competing goals
 such as minimizing development time and  maximizing number of features delivered.
 An example software product line
was the {\bf SCRUM} model shown in Figure~\ref{fig:scrumModel}. Another software product line model  used in this paper represents options
with a {\bf BILLING} system. Apart from that, this paper will also optimize eight other artificially generated SPL models of increasing
size and/or frequency of cross-tree constraints. 

All these software product line models were taken from the SPLOT reseaerch model repository\footnote{http://www.splot-research.org/}.
That website has a tool for artificially generating models. This tool is useful for stress testing an algorithm by e.g., generating models
of increasing complexity. We used eight such models: 
\bi
    \item
    We {\bf increased}    features size while
   using the {\bf same}
      ratio of constraints to features  for
       125FEAT, 250FEAT, 500FEAT. 1000FEAT.
    \item
    In  0.25 C.D, 0.75 C.D, 0.50 C.D and 1.00 C.D,
      the number of features was kept {\bf constant} 
    (at 500)
    while the ratio of constraints to features was 
    {\bf increased}.
\ei
These last ten models
were taken from the SXFM format
and converted into the Dimacs format using FeatureIDE \cite{kastner2009featureide}. And from the Dimacs format, we use a PicoSAT to generate a database of valid solutions for each.

\subsection{Generating Candidates}
This section discusses how we generate candidates
from the models described above.

Chen et al. \cite{chen2018sampling} found that if they generated very large initial populations (e.g., 10,000 candidates), then their
recursive binary chop could very quickly find solutions as good, or better, as genetic algorithms that take an initial population of (say) 100 candidates then mutate them over 100 generations. Following their advice, {\IT}  uses various mechanisms to generate those 10,000 candidates:
\bi 
\item
For models expressed that can be expressed as logical constraints
(e.g., software product lines like
Figure~\ref{fig:scrumModel}) candidate solutions
were generated via PicoSAT v0.6.3\footnote{Which, can be
installed via ``pip3 install pycosat==0.6.3''.}.
\item 
Our other models (specifically XOMO and POM3 variants) came with their own procedural engine for generating examples.
\ei
\subsection{Comparison Algorithms}

Arcuri et al.~\cite{Arcuri11} recommend that algorithms need to be compared against some simple baseline. For that purpose, we use a
Non-Interactive Genetic Algorithm (NGA) which is the same genetic algorithm used by   Araújo et al. but without human interactions. This approach  serves as a baseline for comparison towards answering our second research question, where we only perform y-evaluations to optimize our models.
For the population control parameters of NGA, we used the advice of Holland et al.~\cite{holland92};
i.e., 100 valid candidates and is run for 100 generations, thus generating 100 new candidates each time.
As to the cross-over and mutation method,  we retained the  Araújo et al. settings;
i.e., 90\%
crossover and a  mutation rate set to $1/A$ (where $A$  is the number of attributes in that model).

{\IT} is also compared to a state-of-the-art evolutionary method;
specifically, FLASH~\cite{flash_vivek} and the  Araújo et al. iSBSE~\cite{araujo2017architecture} algorithm
We choose their algorithms for their recency and superior
performance (compared to other methods in their field~\cite{zuluaga2013active}). 
Also, the \cite{araujo2017architecture} algorithm, on the other hand,
 is engineered to accept a wide range of models as input.
 
The code for FLASH, written in Python, comes from
the Nair et al. repository\footnote{ \url{https://github.com/FlashRepo}}.
  Araújo et al.   did not   offer a reproduction package for their work so
 we reimplemented their code from   their descriptions.

\subsection{Manual and Automatic Oracles}\label{evmetrics}

We conducted two studies:
\bi 
\item {\bf Human-in-the-loop}, to see if human  trust or accept   our solutions. 
\item {\bf Fully automated},  to test dozens of models.
\ei
In the former,  humans were  the oracle.
In the latter,  each attribute was given    a  randomly assigned priority (and approved answers  where those using attributes that maximized the sum of that priority).

\subsection{Evaluation Metrics (distance to heaven)}
To evaluate the optimally of  solutions found by different methods, we must see where our solutions fall within the space of all solutions.
To that end we:
\bi 
\item Ran one method (NGA, iSBSE, FLASH, etc) to collect a small number of 
{\bf recommended candidates}.
\item 
Evaluated  {\bf all candidates} and
ranked them all;
\item Find     {\bf recommended candidates}
 ranks for   {\bf all candidates}.
\ei
For ranking {\bf all candidates}, 
using the Ziztler predicate, we can   obtain a vector Z containing all candidates sorted from best to worst. The ``distance to heaven'' of a specific 
{\bf recommended candidate}, denoted $x$
by its index $i_x$ within $Z$ as follows:

        \begin{equation}\label{d2h}
            \mathit{d2h}_{x} = i_x / |Z|
        \end{equation}
Note that this number ranges from 0 to 1 and solutions with {\bf lower} numbers are {\bf better} (since they are closest to 
{\bf heaven}).

 \section{ Results}\label{results}
The results of this paper comment on the research questions offered in the introduction.

\begin{figure}[b!]
    \centering    
    \includegraphics[width=.8\columnwidth]{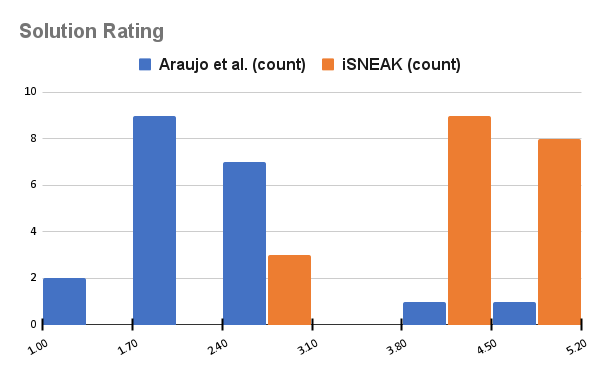}
    
    \caption{RQ1 results. Distribution of ratings given by human experts to presented solutions. The x-axis denotes human ratings and the y-axis shows the frequency counts by which humans offered those ratings.}
    \label{fig:acceptance}
\end{figure}
\subsection{RQ1:  Does {\IT}-style AI confuse humans? }\label{rq1}

{\bf RQ1}  addresses the core issue that motivated this paper;
i.e., {\bf  can we make AI deliver solutions that humans agree with?} 
Stakeholders may have preferences for a tiny
fraction of the total space. Hence, it is  possible that automated
tools like {\IT} will differ from what a stakeholder considers acceptable.
{\bf RQ1} tests for the presence of that problem.

The {\bf RQ1} experiment was performed
in  accordance  with 
the Investigator Review Board policies of the North Carolina State  University.  
Note that in the following experiments, human input is used in all the comparison algorithms studied here (i.e., both {\IT} and  Araújo et al.)
Within the experiment, we only asked humans to compare {\IT} with the iSBSE method proposed by Araújo et al. 
Neither FLASH nor NGA was used here.
We make this choice for two reasons:
\bi
\item
Every added component to a human-in-the-loop study increases the effort associated with our human subjects. Hence we were keen
to minimize their cognitive load.
\item 
FLASH and the NGA behaved so poorly in our automatic trials
that there was no pressing need to test them.
\ei
 To obtain human subjects, we  reached out to our contacts in the Brazilian I.T. community
 where one manager granted us access to 20 of her developers, on the condition we took no more than two hours of their time (for initial briefing and running the experiments). 
Our choice of subjects lead to certain decisions for our experimental design. We had to use a model with attributes that our subjects understood. In consultation with the manager, we reviewed several models (before speaking to subjects) and the decision was that POM3a was the most approachable for our subjects.

For this experiment, subjects were selected by their manager. 
All subjects had at least 4 years of experience in the field and 3 years of experience in an agile team. 
Subjects were made aware that the manager endorsed their participation  in this study.
No added incentives were offered to subjects except a commitment that if in the future they wanted to use the tool, we would make it freely available and support their use.
Subjects and their specific results were guaranteed anonymity from their manager. Thus, the experiment did not collect logins, names, or IP addresses.

\begin{figure}[!b]
    \centering
    \includegraphics[width=.8\columnwidth]{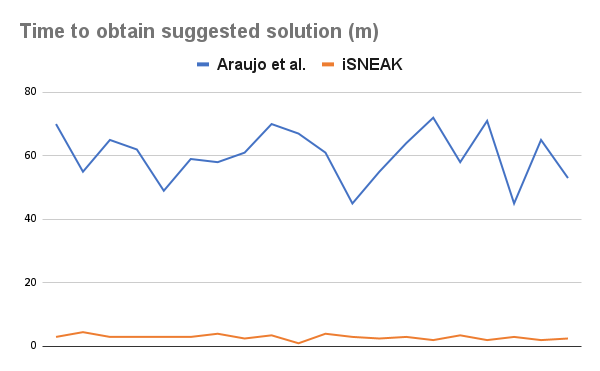}
    \caption{RQ1 results: Human effort required for each solution, in minutes }
    \label{fig:humaneffort}
\end{figure}

Prior to collecting data, we ran a short (under 20 minutes) in-person meeting with    participants (including the manager of the group, who also was as a subject).
Subjects were introduced to the goals of the work and   briefed on the models. After that, subjects had 1.5 hours to complete the experiment during which time, they were asked not to talk to each other about their experiences. 
During    the experiment,   subjects were silently observed in a company office by one of the authors. 
Subjects installed and ran our software locally on their own machines.
At the end of the experiment, subjects rated two configurations (using a score of 0 to 5, 0 being the worst).

In order to control for experimental-subject bias, after running both the
Araújo et al.'s tool and {\IT} tools,   solutions were presented in a randomized order (so that the {\IT} solutions did not always appear in the same place on the screen).
This meant that our subjects never knew from which tool came each solution they were asked to evaluate.

As seen in Fig.~\ref{fig:acceptance},  our subjects strongly   preferred {\IT}.
The median score for {\IT}'s solutions was 4
while
the median score for 
Ara\'ujo et al.'s solutions was 2.
Hence for RQ1, we say: 
\begin{formal}
For models we could show to our subjects in their available time, SNEAK's solutions are acceptable.
\end{formal}

We have   recorded the time it took each participant to obtain a solution with each tool. As seen in Fig.~\ref{fig:humaneffort}, {\IT} requires orders of magnitude less human time   in order to obtain preferable solutions.

\subsection{RQ2: Does human advice confuse iSNEAK-style AI?}  \label{rq2}

  {\bf RQ2} addresses the following question: {\bf are human opinions  detrimental to optimization}, or {\bf if we generate solutions using the methods of RQ1,  will we achieve sub-optimal results?} 

To answer this question, we explore
numerous models other than the POM3a model
used for {\bf RQ1}. 
To facilitate this process, we used some   oracle that can comment on 320 results; i.e., 20 repeats over 16 models (many of which are unfamiliar to specific subject matter experts). Hence we needed to build an automatic oracle. 

To do this, 
before any interaction in each run,
our  oracle would randomly assign priority values for all of the model's variables. Then using those priority values it would be able to consistently respond to any oracle questions posed by our algorithms.
These priority values are set according to a random seed.

\begin{figure}[!t]
 \caption{RQ2 results. Median {\bf d2h} values seen over 20 runs (Log scale on the Y axis). For a definition of {\bf d2h}, see Equation~\ref{d2h} in \ref{evmetrics}. Note that \underline{{\bf lower}} values are \underline{{\bf better}}.}
    \label{fig:d2h}
 \centering
    \includegraphics[width=1\linewidth]{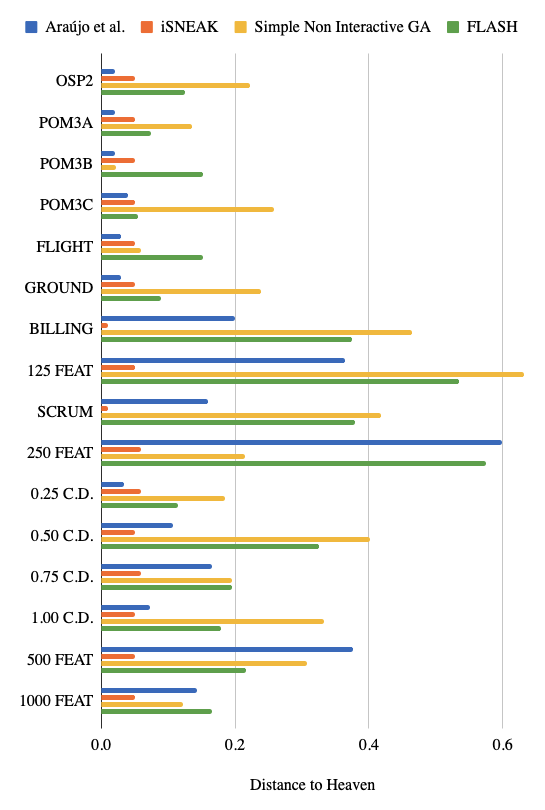} 
\end{figure} 

\begin{figure}
 \caption{RQ2 results.
    Any iSBSE tool will, $I$ times,
    ask a stakeholder about $S$ attribute.  Shown here are the 
    $I,S$ numbers seen in 20 runs
    (different random number seeds each time).  }
    \label{fig:Interaction}
     \vspace{1cm}
    \centering
    \includegraphics[width=.8\linewidth]{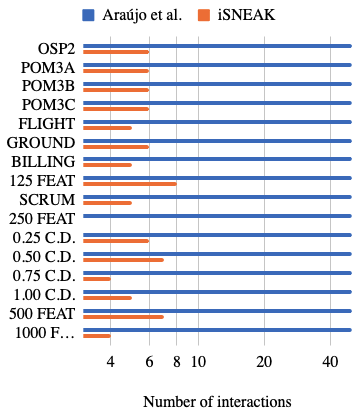}\\ 
    {\scriptsize \bf Fig.~\ref{fig:Interaction}a :    Number of interactions $I$ (medians from 20 runs).
   Here,  \underline{{\bf lower}} values are \underline{{\bf better}} (logarithimic scale). Note that the blue bars have a fixed top value since that parameter is hard-wired into the Ara\'ujo et al. architecture.}

    \vspace{1cm}
    
     \includegraphics[width=.8\linewidth]{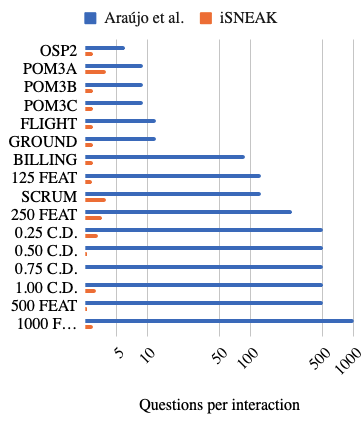}\\
     {\scriptsize \bf Fig.~\ref{fig:Interaction}b:  Median questions $S$ (in 20 repeats) asked per interaction.  Here,
     \underline{{\bf lower}} values are 
     \underline{\bf better} (logarithimic scale).} 
    
\end{figure}

All results come from 20 repeated runs (with different random number seeds) of an automatic oracle exploration of our models.

  Fig.~\ref{fig:d2h} comments on the effectiveness of {\IT}'s questions. Usually,
  {\IT} can find results within 2\% to 3\% of the best solution ever seen for a model. More specifically, in Fig.~\ref{fig:d2h}:

\bi
\item
For all models, the simple GA usually performs worst.
 
\item FLASH always ranks third or fourth  best.
\item
 Araújo et al.'s tools have   medians    better than {\IT} for  \\ unconstrained models (the first 6 models). But for models with constraints (i.e., from BILLING downwards), {\IT} out-performs Araújo et al.
Sometimes, those wins are very significant:
e.g.,  see BILLING here {\IT}'s median {\bf d2h} scores are an order of magnitude better than the other algorithms.
\item
When {\IT} lost to Araújo et al., the difference is usually very small
(the actual {\bf d2h} performance deltas from the best results are $\{1,2,2,3,3,3\}\%$).
\ei
 Fig.~\ref{fig:Interaction}a shows that when Araújo et al. y-evaluates its 10,000 candidates, it pauses around 50 times to ask stakeholders some questions. As shown in   Fig.~\ref{fig:Interaction}b, these questions cover all the attributes in the model, which means the size of the interaction for Araújo et al. scales linearly with the number of variables in the model.
 On the other hand, {\IT}  pauses around 20 times, and when it does, it asks far fewer questions, and as seen in Fig.~\ref{fig:Interaction}b the size of {\IT}'s interactions has no direct relationship with the size of the model. 
 
Just to clarify, the reason  Araújo et al. asks more questions than {\IT} is that 
when  Araújo et al.,  ask the stakeholders questions, those questions  mention {\bf every} attribute in the model (e.g., all 128 attributes of Figure~\ref{fig:scrumModel}).
On the other hand, before {\IT} asks a question, it applies   the attribute selection methods of \ref{isneak}
in order to isolate the most informative attributes.

We answer   {\bf RQ2} as follows:
\begin{formal}
For large and complex models with many constraints,
{\IT} is the clear winner  (of the systems studied here).
\end{formal}
To be clear, 
sometimes {\IT} is defeated by other methods-- but only by a very small amount and only
in  simple models (that are both very small and have no constraints).
More importantly, measured by the human cost to find a solution, {\IT} is clearly the preferred method.   We say  cost
is {\bf number of interactions} times {\bf number of questions per interaction}. 
Using Fig.~\ref{fig:d2h}, we compare the cost 
for our different systems. {\IT}'s oracle cost was    1\%,4\% of   Araújo et al.  
for the constrained and unconstrained models

 \begin{figure}[!t]
    \centering
    \includegraphics[width=.8\columnwidth]{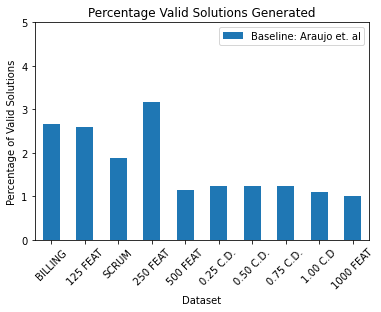}
    \caption{Valid Solutions over 20 runs (Technical aside: to handle these invalid solutions,  we added
            a post-processor to the 
            Araújo et al. algorithm that rejected
            invalid solutions.)}
    \label{fig:ValidPercentage}
\end{figure}

\subsection{Other Issues}

We have other reasons to recommend {\IT} over the methods of 
Araújo et al.: 
 100\% of all the solutions
explored by {\IT} are valid. The reason for this
is simple: we let PicoSAT, or the model generative tool, generate valid solutions,
then we down-sample from that space. Genetic algorithms, on the hand, such as those used in Araújo et al. mutate
examples without consideration of the logical constraints of a domain.

This is a significant and important aspect of {\IT}.
Fig.~\ref{fig:ValidPercentage}
showed what happens when we take the solutions
generated by Araújo et al.'s genetic algorithm~\cite{araujo2017architecture} for the SPL models, then applied
the model constraints to those solutions. As seen
in that figure, most of the solutions generated 
by their methods are not valid. This also applies to NGA and FLASH, which waste most of their time evaluating invalid solutions.

\section{Discussion}
\subsection{Threats to Validity}\label{threats}

As with any empirical study, biases can affect the final
results. Therefore, any conclusions made from this work must
be considered with the following issues in mind.

\noindent\textbf{Evaluation Bias} - In our {\bf RQ1} experiments, we studied   20 software engineers using {\IT} to select for a good POM3A \cite{boehm2003using, boehm2003balancing, port2008using} solution (a model for exploring the management process of agile development). Although the number of professionals could be higher to provide us with a more reliable evaluation of the approach, we have mitigated this by selecting only professionals with at least 4 years of experience in the field and 3 years of experience in being in an agile team. 

\textbf{Parameter bias} - {\IT} contains many modular subprocesses within itself, containing different  hyperparameters. 
While this study shows that the current configuration of {\IT} is capable of producing results comparable to state-of-the-art algorithms, different configurations of the {\IT} system might prove valuable. 
That said, in defense of our current setup, we were able to find solutions that were significantly better than those found by the compared baseline techniques.

\noindent\textbf{Sampling Bias} - This threatens any empirical study using datasets. i.e, what works here may not work everywhere. For example, we have applied {\IT} to two real-life models of different software product lines, six XOMO and POM3 models, and eight artificially  generated feature models of varying characteristics. But, the behavior of {\IT} on significantly larger models (i.e., hundreds of thousands of attributes) still needs to be evaluated.

Another concern is that if the models studied here are ``trivial''
in some sense, then there may be very little value added by {\IT}. We believe these these models are not trivial for several reasons.
In fact,  for all practical purposes, these models have a  search space so
large that  it cannot be enumerated. Consider, for example, the SCRUM model with its 128 binary options. From Figure~\ref{fig:ValidPercentage}, we see that (usually) 98\% of thee randomly generated solutions might violate domain constraints. 
That still leaves a space for at  least $2^{128}*0.02=10^{38}$ solutions. 
In our experience with PicoSAT, it 
takes 12 hours to generate 100 million solutions\footnote{Since every time we find a new solution, we stop PicoSAT from finding it again. This requires negating
the latest solution, adding it back into the models, the re-running the algorithm from scratch.}.
Hence  $2^{128}*0.02$ solutions would require nearly $10^{26}$ years
to enumerate.

\noindent\textbf{Algorithm bias}
Our reasons for using the
the   \cite{araujo2017architecture} algorithm were described in \ref{isbse}, but future work needs to explore other algorithms. 
There are many other state-of-the-art optimizers such as OPTUNA~\cite{akiba2019optuna} and HYPEROPT~\cite{bergstra2013hyperopt}
and many more aside.

\subsection{Relationship to Prior Work}

As to our own prior work with iSNEAK~\cite{lustosa2024learning}, we applied an earlier version
of this algorithm   to the problem of learning
from very small data sets.
This paper differs from that prior work in several significant ways.
That article  
 lacked any facility for  human-in-the-loop interaction. 
 But here, as discussed
in \ref{evmetrics}, we exploit that facility  in two 
important ways:
\bi 
\item In one of our studies,    we  ask humans for their opinions during the  optimization process
(and those opinions are used to guide the results).  
\item 
In another of our studies, that explores many  models, we guide the inference with  an ``artificial human'' with their own built-in bias (that we create as part of this study, see \ref{evmetrics}).
\ei 
Aslo, that  prior article was a hyperparameter optimization study where
  a SNEAK-like algorithm was used to control a second algorithm (a Random Forest predictor). Here, there is  no second algorithm (so the output of iSNEAK is the final output).

Further, that prior article was focused on a highly specific 
``small data'' problem:   tuning learners to explore tiny data sets (60 rows or less). 
This article, on the other hand, is a ``big data'' study that explores much bigger problems:
\bi
\item That article explored data with just five independent variables while here, we explore data sets with up to 1000  independent variables.
\item That article explored data data sets with 60 rows (or less) while this article explores samples of model output of size 10,000.
\ei

\subsection{Future Work}
 
Going forward, these results needs to be applied to more models.
Also, we need more results from more human-in-the-loop studies.
Further, here we found that the {\IT} pre-processor helped one particular optimizer (SWAY) 
and it could be insightul to check if this pre-processor is helpful for other optimizers.

Furthermore, \cite{ochoa2018systematic}  propose   a set of challenges that need to be addressed for semi-automatic configuration  tools  including (a)~tuning all the control parameters of tools
in {\IT}; (b)~supporting multi-stakeholder environments; (c)~reasoning about qualitative requirements (such as the non-functional requirements explored by \cite{mathew2017short}); (d)~real-time perpetual re-evaluation of solutions in highly dynamic environments (e.g., drones proving assistance within   disaster sites); and (e)~exploring different algorithms. 

As to this last point (exploring different algorithms), we note that this work used a particular clustering algorithm (FASTMAP) and a particular
attribute selector (INFOGAIN). Clearly, future work should explore  better clustering and attribute selection algorithms. 

More generally, there is an as-yet unexplored connection between
our work here and semi-supervised learning (SSL).
Given a few  evaluated examples, SSL tries to spread out those labels across related areas in the data set~\cite{liu2016comparative,mit06}. 
A key assumption of SSL is that higher-dimensional data sets can be approximated by a lower-dimensional manifold, without little to no loss of signal~\cite{mit06}. When such manifolds exist, then  the number of queries required to understand   data is just  the number of queries needed to understand the manifold.  A common way to find that manifold is to apply some dimensionality reduction method such as PCA. 
We saw in  \ref{isneak}  that  {\IT} uses an analog for PCA. Hence,  in some sense it could be said that {\IT} is a semi-supervised learner. 
That said, we have yet to find an algorithm from the SSL literature that improves on the results of \ref{rq2}. Nevertheless, this could be a fruitful direction for future study; i.e.
\bi  
\item For all the SSL algorithms,
see if any of them do better than {\IT}.
\ei

\section{Conclusion}
 
When   human  knowledge falls within a small fraction of the total
  space then it is vanishingly unlikely that a fully automated algorithm will select   solutions that
 are comprehensible and acceptable. However, as our {\bf RQ1} results show, it is possible
 to select solutions that make sense to human subject matter experts.

Also, when humans guide the reasoning, but can only comment on a tiny fraction of the total problem,
this   limited knowledge might lead to sub-optimial results. However, as shown by our {\bf RQ2} results,   it is possible for AI tools to respect human preferences while still delivering highly optimal solutions.
 
The key to this process are quickly computed ``hints'' that produce approximate partial
orderings of the data. 
In our first round of ``hinting'',   {\IT}   partially orders,  
then prunes, recursive partitions of the data (using questions to some oracle about the independent attributes).
Next, in a second round of ``hinting'' we used  Equation~\ref{eqzitzler}
to      order,  then prune the data (using queries to the goal attributes).

This paper tests this process on numerous software models.
For the
models studied here:
\bi 
\item  iSNEAK-style AI does not confuse humans since
it only asks a handful of questions and returns human-
acceptable solutions,
\item 
For the models studied here, advice from humans do
not confuse AI since iSNEAK’s solutions, obtained via
human-in-the-loop reasoning out-perform the prior state-
of-the-art in the optimization of software models.
\ei

\section*{Acknowledgments}
This work was partially supported by an 
NSF CCF award \#1908762
 Our human experiments were performed in accordance 
  with the
 Investigator Review Board policies of the
North Carolina State University, IRB protocol \#24233.

\section*{Conflict of Interest Statement}
The authors declared that they have no conflict of interest.
\section*{Data Availability Statement}
All our data and scripts are on-line at
\url{https://github.com/zxcv123456qwe/iSNEAK}.
Permission is  granted, free of charge, to any person obtaining a copy
of this software and associated documentation files (the "Software"), to deal
in the Software without restriction, including without limitation the rights
to use, copy, modify, merge, publish, distribute, sublicense, and/or sell
copies of the Software, and to permit persons to whom the Software is
furnished to do so, subject to the   
conditions of the MIT License.

\bibliographystyle{IEEEtran}
\bibliography{bibliography,other,bibs/kenrefs,bibs/proposal,bibs/ssl,bibs/suvodeep}

\begin{IEEEbiography}[{\includegraphics[width=1in,height=1.25in,clip,keepaspectratio]{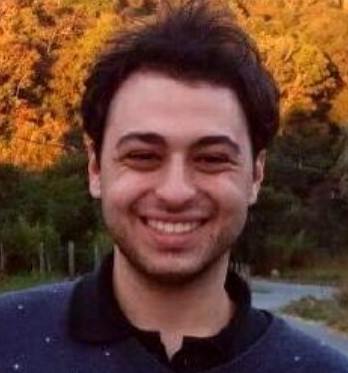}}]{Andre Lustosa} is a fifth year PhD student in Computer Science at NC State University. His research interests include interactive search based software engineering, machine learning for software engineering, and landscape analysis for software analytics.
\end{IEEEbiography}

\begin{IEEEbiography}[{\includegraphics[width=1in,height=1.25in,clip,keepaspectratio]{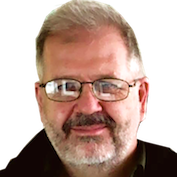}}]{Tim Menzies} (IEEE Fellow, Ph.D. UNSW, 1995)
is a Professor in computer science  at NC State University, USA,  
where he teaches software engineering,
automated software engineering,
and programming languages.
His research interests include software engineering (SE), data mining, artificial intelligence, and search-based SE, open access science. 
For more information,  please visit \url{http://timm.fyi}.
\end{IEEEbiography}

\EOD
\end{document}